**Coronagraph Experiment on Dark-hole Control by Speckle Area Nulling Method**
Masahito Oya[1,2*], Jun Nishikawa[3,4], Masaaki Horie[1], Kazuma Sato[5,2], Naoshi Murakami[6], Takayuki Kotani[3], Shiomi Kumagai[7], Motohide Tamura[8,3], Yosuke Tanaka[9], Takashi Kurokawa[3,9]

[1] *Graduate School of Science and Technology, Nihon University, Kandasurugadai 1-8-14, Chiyoda, Tokyo 101-8308, Japan*
[2] *Division of Optical and Infrared Astronomy, National Astronomical Observatory of Japan, Osawa 2-21-1, Mitaka, Tokyo 181-8588, Japan*
[3] *Extra-Solar Planet Office, National Astronomical Observatory of Japan, Osawa 2-21-1, Mitaka, Tokyo 181-8588, Japan*
[4] *School of Physical Sciences, The Graduate University for Advanced Studies, Osawa 2-21-1, Mitaka, Tokyo 181-8588, Japan*
[5] *Graduate School of Engineering, Tokyo University of Agriculture and Technology, 2-24-16 Naka-cho, Koganei, Tokyo 184-8588, Japan;*
[6] *Facility of Engineering, Hokkaido University, Kita-13, Nishi-8, Kita-ku, Sapporo, Hokkaido 060-8588, Japan*
[7] *Facility of Science and Technology, Nihon University, Kandasurugadai 1-8-14, Chiyoda, Tokyo 101-8308, Japan*
[8] *Faculty of Science, University of Tokyo, 7-3-1, Hongo, Bunkyo-ku, Tokyo 113-0033, Japan*
[9] *Faculty of Engineering, Tokyo University of Agricalture and Technology, Koganei, Tokyo 184-8588, Japan*

**Abstract**
In high-contrast imaging optical systems for direct observation of planets outside our solar system, adaptive optics with an accuracy of λ/10,000 root mean square is required to reduce the speckle noise down to $1 \times 10^{-10}$ level in addition to the nulling coronagraph which eliminate the diffracted light. We developed the speckle area nulling (SAN) method as a new dark-hole control algorithm which is capable of controlling speckle electric field in a wide area quickly, in spite of an extension of speckle nulling, and is robust not relying upon an optical model. We conducted a validation experiment for the SAN method with a monochromatic light and succeeded in reducing the intensity of areal speckles by 4.4× $10^{-2}$.

1. **Introduction**

Various high-contrast imaging optical systems are being developed for direct observation of planets outside our solar system. For direct observations, particularly of terrestrial planets, a contrast ratio between a star and planets of eight to ten orders of magnitude should be overcome. A nulling coronagraph [1-5] is employed to eliminate the diffracted light of the aplanatic wavefront for that purpose. Furthermore an adaptive optics with an accuracy of λ/10,000 root mean square (rms) must be implemented even in a space telescope with a stable wavefront, in order to reduce the intensity of remaining speckles in the image plane produced by aberrations of optics, where λ is the wavelength of the light.

Since it is difficult to achieve such a high accuracy with ordinary adaptive optics, a "dark-hole" control method is often used, in which a deformable mirror (DM) is operated to make the speckles in a specific region in the final image plane disappear, i.e., creating a dark hole. The speckle nulling (SN) [6-8] and electric field conjugation (EFC) methods [9,10] have been developed for the dark-hole control. The SN method makes a sequential extinction of speckles starting from the brightest point. The drawback of this method is that the rate of extinction for a wide area is slow. Meanwhile, the EFC method can reduce an intensity of a wide area in one lump, but this requires responses of the real and imaginary parts of the electric field (EF) in the image plane to variations of the DM actuator, which would include the coronagraph optics, be modeled in advance. The method, therefore, depends on the response model. In order to overcome these drawbacks, we developed the speckle area nulling (SAN) method as a new dark-hole control algorithm. The SAN method is capable of controlling speckle EFs of a wide area quickly, in spite of an extension of SN, and robust not relying upon a response model. This paper will describe an experiment of the SAN method.

2. **Principle of Speckle Area Nulling (SAN) Method**

A brief explanation of the principles of the SAN method is shown in this section. First, incoming

E-mail address: masahito.ooya@nao.ac.jp

starlight is guided to the high-contrast optical system after the focal point of a telescope. The diffracted light is eliminated by a coronagraph in the system, and the intensity of speckles is reduced by the SAN method, which uses the DM on the pupil plane, as well as the camera at the final image plane (Fig. 1).

In a typical application of the traditional SN method, a wavefront of sinusoidal shape is generated with a DM. By the relationship of Fourier transform, the sinusoidal wavefront then produces a modulated EF on the final image plane that has two peaks point-symmetrical to the optical axis and has a distribution pattern convolved by point spread function (PSF). Here, the PSF is derived through a Fourier transform of the aplanatic EF of the pupil plane. It becomes the well-known Airy pattern in the case of a circular aperture. The amplitude, argument, and position of the modulated EF can be controlled by the amplitude, spatial phase, and wave number of the sinusoid, respectively. This modulated EF is used to take measurements of and eliminate the speckle EF. It was possible to extinguish multiple speckles simultaneously by overlapping sinusoids with different wavenumbers, although the speckles must be separated farther than about $3\lambda/D$ not to diverge the control affected by the envelopes of the PSF, where $\lambda/D$ is an angular unit determined by $\lambda$ and the diameter D of the pupil (telescope). Therefore the convergence to the dark hole was slow.

On the other hand, in the SAN method, all of the pixels closer than $1\lambda/D$ in the target area are intended to apply the same control as the SN method. In order to secure the spatial resolution for astronomical observations, the pixel interval is set to a fraction of the full width at half maximum (FWHM) of the PSF. This leads to the modulated EF being the sum of overlapped PSF generated by the sum of sinusoidal wavefront. In this condition of the SAN method, however, it was a discovery that the control did show not divergence but rather fast conversion to the dark hole. We describe the sum of the sinusoids later, show how to derive a relationship between the modulated EF and the speckle EF from measured intensities first.

Let us consider the speckle EF denoted as $E_0$ at the $i$-th ($i=1, 2, \cdots, N$) pixel of the camera in the target dark-hole region, and then the intensity $I_0 = |E_0|^2$, where $E_0$ is a complex number. For the purpose of simplicity, the subscript $i$ indicating the pixel number for $E_0$ and $I_0$ is omitted. The DM mounted on the pupil plane is used to modulate the wavefront phase into four sinusoidal types, namely those of $\pm\sin$ and $\pm\cos$, to modulate the final image plane into four types of EFs, namely $\pm\Delta E_1$ and $\pm\Delta E_2$, respectively (Fig. 2). The four intensities that correspond to these conditions, namely $I_1^+, I_1^-, I_2^+$, and $I_2^-$, respectively, are

$$\begin{cases} I_1^+ = |E_0 + \Delta E_1|^2 \\ I_1^- = |E_0 - \Delta E_1|^2 \\ I_2^+ = |E_0 + \Delta E_2|^2 \\ I_2^- = |E_0 - \Delta E_2|^2 \,. \end{cases} \quad (1)$$

Here, the subscript $i$ is omitted for $\Delta E_1, \Delta E_2, I_1^+, I_1^-, I_2^+$, and $I_2^-$, as well. Then we obtain following relations

$$\begin{cases} \Delta E_1 \cdot E_0 = (I_1^+ - I_1^-)/4 \\ \Delta E_2 \cdot E_0 = (I_2^+ - I_2^-)/4 \\ |\Delta E_1|^2 = (I_1^+ + I_1^- - 2I_0)/2 \\ |\Delta E_2|^2 = (I_2^+ + I_2^- - 2I_0)/2 \,. \end{cases} \quad (2)$$

Here, $\cdot$ is the inner product when the EF is considered as a two-dimensional vector in a Cartesian coordinate. Since the modulated EFs, $\Delta E_1$ and $\Delta E_2$, are perpendicular to each other in the complex plane, the speckle EF can be written as

$$E_0 = \frac{\Delta E_1 \cdot E_0}{|\Delta E_1|^2}\Delta E_1 + \frac{\Delta E_2 \cdot E_0}{|\Delta E_2|^2}\Delta E_2 \,. \quad (3)$$

Therefore, based on equation (2) and (3), the speckle EF is

$$\begin{cases} E_0 = p\Delta E_1 + q\Delta E_2 \\ p = (I_1^+ - I_1^-)/2(I_1^+ + I_1^- - 2I_0) \\ q = (I_2^+ - I_2^-)/2(I_2^+ + I_2^- - 2I_0) \,. \end{cases} \quad (4)$$

Here, the subscript $i$ is omitted for $p$ and $q$, as well.

Next we should describe the phase distribution in the pupil plane for the modulation. In experiments, however, a voltage map applied onto the DM in the pupil plane is convenient which is proportional to the phase including a response of the DM, and can be used directly for the control. The modulation voltages applied on the respective actuators to obtain the modulated EFs, $\Delta E_1$ and $\Delta E_2$, can be written as

$$\begin{cases} V_1^+(\xi,\eta) = \sum_i V_0 \sin\left\{\frac{2\pi}{\lambda f}(x\xi + y\eta)\right\} \\ V_2^+(\xi,\eta) = \sum_i V_0 \cos\left\{\frac{2\pi}{\lambda f}(x\xi + y\eta)\right\} \end{cases}, \quad (5)$$

respectively, where $(\xi,\eta)$ indicates the position of the actuator of the DM, $(x,y)$ is the coordinate for the $i$-th pixel, $f$ is focal length of the imaging system, $V_0$ is an amplitude of voltage for each sinusoidal modulation. Here, the subscript $i$ is omitted for $x$ and $y$, as well. In the experiments, $V_0$ was set so as to make the maximum of the modulated intensities twice of the maximum unmodulated intensity among the target N pixels, where a transmittance of the sinusoidal EF at the coronagraph optics was included. By using equations (4) and (5), the SAN method adopt the control voltage written as

$$V(\xi,\eta) = -\sum_i \left[ pV_0 \sin\left\{\frac{2\pi}{\lambda f}(x\xi + y\eta)\right\} + qV_0 \cos\left\{\frac{2\pi}{\lambda f}(x\xi + y\eta)\right\} \right] \quad (6)$$

for producing $-(p\Delta E_1 + q\Delta E_2)$ and removing the speckle EF of the target N pixels. Thus the speckle EFs of a wide area can be controlled simultaneously not relying upon any model.

## 3.    Experimental Result

An explanation of the experiment on the SAN method that used the vector vortex coronagraph (VVC) [11,12] is shown in this section. In the laboratory, the 671 nm DPSS laser beam guided by a single mode fiber was collimated and passed through the entrance pupil, as shown in Fig. 1, and the image was formed to simulate starlight arriving at the focal point of the telescope. The beam was then fed through the DM and the VVC and came at the CCD camera, where the final focal image was measured. The VVC eliminated diffracted light at the Lyot stop by using the vortex mask to generate a phase difference that was proportional to the azimuth angle at a position of the image plane ($4\pi$ for one round or the like). In our experiment, a VVC capable of eliminating light across a wide band was constructed by installing an axially-symmetric half-wave plate at the focal point and sandwiching it with quarter wave plates and linear polarizers aligned in a crossed arrangement [12]. The DMs, with dimensions of 12 x 12 manufactured by the Boston Micromachines Corporation were used at the pupil plane for the control with the SAN method. An actuator pitch of the DM was 0.30mm and the pupil diameter of the optics was 3.0mm. The experimental results by performing controls repeatedly are shown in Fig. 3, where a single control consisted of a set of five measurements and a correction. Since the outermost boundary of the target region were limited to about $5\lambda/D$ by the Nyquist frequency of the actuator pitch of the DM. A single pixel of the camera on the image plane was $0.19\lambda/D$ in the experiment. The innermost boundary of the target region was carefully set in order to perform stable control. Since there was only one DM and only the phase of the wavefront was being controlled, the target region was only on the half side of the star. This was because the two centrosymmetric modulated EF by sinusoidal wavefront had a relationship being negative of the complex conjugate, both could not be eliminated simultaneously. Fig. 3(a) shows an image of the final image plane, prior to controls. The region surrounded by the half-ring was the target region (a region delineated by $\xi < -0.19\lambda/D$ and $0.97\lambda/D \leq \rho \leq 4.4\lambda/D$, where $\rho = \sqrt{\xi^2 + \eta^2}$. The speckle intensities of the target region prior to controls (Fig. 3(b)) were approximately $1 \times 10^{-4}$ to $1 \times 10^{-5}$. Controls were implemented 12 times and resulted in a successful areal extinction as shown in Fig. 3(c), and the average intensity of this area was $1.4 \times 10^{-6}$. This corresponds to an extinction of $4.4 \times 10^{-2}$ from the initial value of $3.1 \times 10^{-5}$. The average extinction for each implementation is shown in Fig. 3(d). The first few controls resulted in significant reductions, after which gradual saturation was observed. It matched with a simulation result which relative extinction of $1.2 \times 10^{-2}$ by 12 times implementations as shown in Fig. 3(d), too, with an initial wavefront error of $\lambda/60$ rms and amplitude error of 0.030 rms. When we evaluate a narrow half-moon-shaped region of $\xi \leq -1.5\lambda/D$ and $\rho \leq 4.4\lambda/D$, the average contrast of this area reached $5.3 \times 10^{-7}$. A simulation showed that a relative extinction better than $1 \times 10^{-3}$ would be possible if wider regions could be controlled by a DM with larger number of actuators.

## 4.    Summary and Conclusion

The SAN method is a new dark-hole control algorithm that takes measurements of and eliminates the speckles areally for high-contrast imaging optical systems having a coronagraph and an adaptive optics. A primary advantage of the SAN method is the quick areal extinction, as well as the robustness not relying upon any model calculations of the optical system. We conducted a validation experiment for the SAN method with a monochromatic light and succeeded in reducing the intensity of areal speckles.

**Acknowledgements**
A part of this work was supported by Grants-in-Aid (No.24360029) from the MEXT. This experiment was being performed in the optical experiment facility of Advanced Technology Center of NAOJ.
**References**
1) M. J. Kuchner and W. A. Traub: Astrophys. J. **570** (2002) 900.
2) J. Trauger, D. Moody, B. Gordon, J. Krist and D. Mawet: Proc. SPIE **8442** (2012) 8442Q.
3) D. Mawet, E. Serabyn, D. Moody, B. Kern, A. Niessner, A. Kuhnert, D. Shemo, R. Chipman, S. McClain, and J. Trauger: Proc. SPIE **8151** (2011) 81511D.
4) N. Murakami, J. Nishikawa, W. A. Traub, D. Mawet, D. C. Moody, B. D. Kern, J. T. Trauger, E. Serabyn, S. Hamaguchi, F. Oshiyama, M. Sakamoto, A. Ise, K. Oka, N. Baba, H. Murakami and M. Tamura: Proc. SPIE **8442** (2012) 844205.
5) M. Mas, P. Baudoz, R.Galisher and G. Rousset: Proc. SPIE **8446** (2012) 844689
6) F. Malbet, J. W. YU, M. Shao: Publ. Astron. Soc. Pacific **107** (1995) 386.
7) J. T. Trauger, C. Burrows, B. Gordon, J. J. Green, A. E. Lowman, D. Moody, A. F. Niessner, F. Shi, and D. Wilson: Proc. SPIE **5487** (2004) 1330.
8) P. J. Borde, W. A. Traub: Astrophys. J. **638** (2006) 488.
9) J. Trauger and W. A. Traub: Nature **446** (2007) 771.
10) A. Give'on, B. D. Kerna, and S. Shaklana: Proc. SPIE **8151** (2011) 85110.
11) D. Mawet, P. Riaud, J. Surdej, and O. Absil: Astrophys. J. **633** (2005) 1191.
12) N. Murakami, S. Hamaguchi, M. Sakamoto, R. Fukumoto, A. Ise, K. Oka, N. Baba and M. Tamura, Opt. Express **21** (2013) 7400.

Figure Caption

Fig. 1 Optical layout

Fig. 2 Modulation of speckle EF

Fig. 3 CCD images and average contrast
(a) Before correction. (b) Enlarged image of before correction. Standardization was performed with the intensity of the off-axis light. The average contrast was about $3.1\times10^{-5}$. (c)After correction. The average contrast was about $1.4\times10^{-6}$. (d) A log plot of contrast of average intensity vs. number of iterative controls in the target region for the experiment and a simulation.

Fig.1

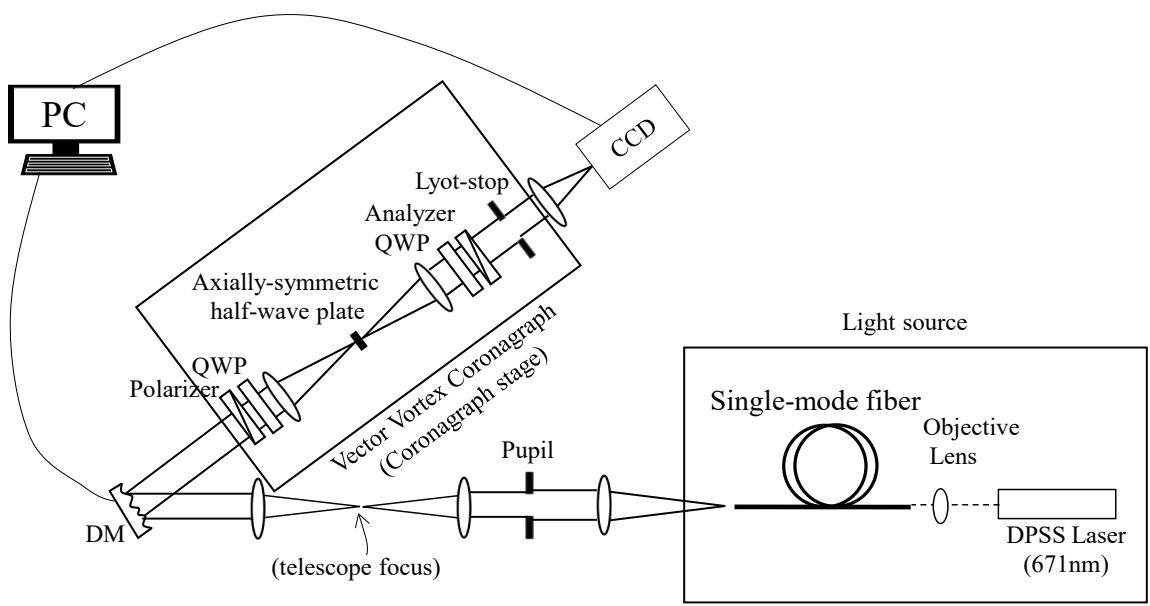

Fig.2

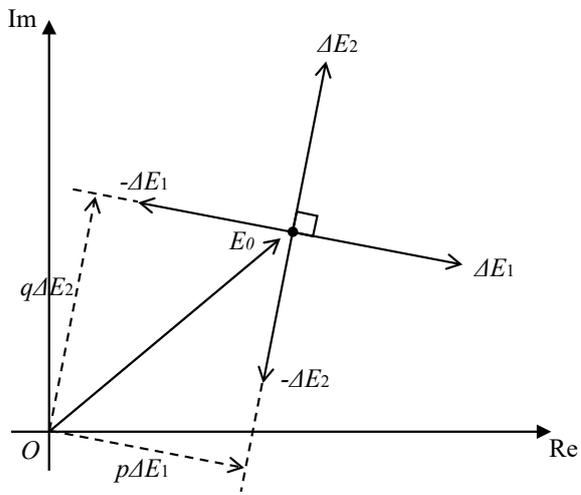

Fig.3

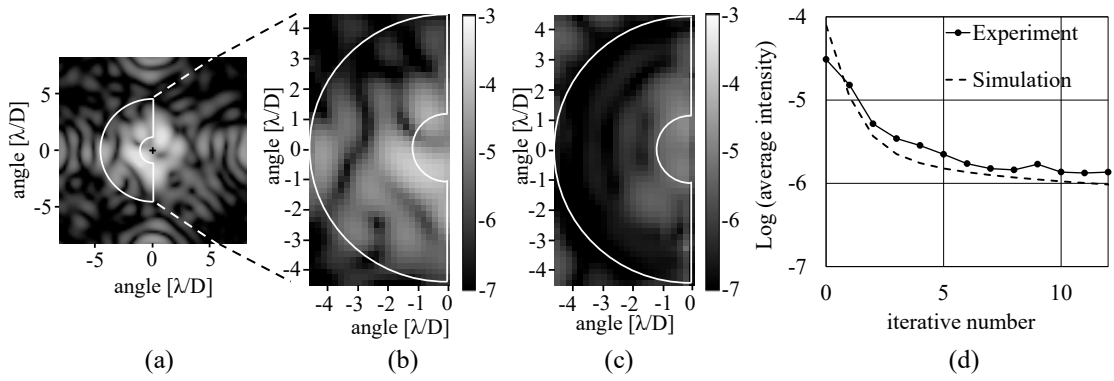

(a)   (b)   (c)   (d)